\documentclass[letter,traditabstract]{aa}
\usepackage{graphicx,natbib}
\usepackage{txfonts}
\usepackage{psfig}

\newcommand{\ltsima} {$\; \buildrel < \over \sim \;$}  
\newcommand{\gtsima} {$\; \buildrel > \over \sim \;$}  
\newcommand{\lta} {\lower.5ex\hbox{\ltsima}}  
\newcommand{\gta} {\lower.5ex\hbox{\gtsima}}  
\newcommand{\Ha} {H$\alpha$}  
\newcommand{\Hb} {H$\beta$}  

\newcommand{\ergsHz}{\ensuremath{\mathrm{\,erg}{\mathrm{\,s}^{-1}{\mathrm{\,Hz}^{-1}}}}}
\newcommand{\ergs}{\ensuremath{\mathrm{\,erg}{\mathrm{\,s}^{-1}}}}

\newcommand{\forb}[2]{\mbox{$[{\rm #1\, #2}]$}}
\newcommand{\oiii}{\forb{O}{III}}
\newcommand{\oiiihb}{\oiii/\Hb}

\begin{document}
\title{An optical spectroscopic survey of the 3CR sample of radio galaxies
  with z $<$ 0.3.}

\subtitle{IV. Discovery of the new spectroscopic class of relic radio
  galaxies\thanks{Based on observations made with the Italian Telescopio
    Nazionale Galileo operated on the island of La Palma by the Centro Galileo
    Galilei of INAF (Istituto Nazionale di Astrofisica) at the Spanish
    Observatorio del Roque del los Muchachos of the Instituto de Astrofisica
    de Canarias.}}

\author{Alessandro Capetti \inst{1} \and Sara Buttiglione \inst{2} \and David
  J. Axon \inst{3,4} \and Andrew Robinson \inst{4} \and Annalisa Celotti
  \inst{5} \and Ranieri D. Baldi \inst{1,5} \and Marco Chiaberge \inst{6,7}
  \and F. Duccio Macchetto \inst{6} \and William B. Sparks \inst{6}}
   
\offprints{A. Capetti}
     
\institute{INAF - Osservatorio Astronomico di Torino, Strada Osservatorio 20,
  I-10025 Pino Torinese, Italy \and INAF, Osservatorio Astronomico di Padova,
  Vicolo dell'Osservatorio 5, I-35122 Padova, Italy \and School of
  Mathematical and Physical Sciences, University of Sussex, Falmer, Brighton
  BN1 9RH, UK \and Department of Physics, Rochester Institute of Technology,
  85 Lomb Memorial Drive, Rochester, NY 14623 \and SISSA-ISAS, Via Beirut 2-4,
  I-34014 Trieste, Italy \and Space Telescope Science Institute, 3700 San
  Martin Drive, Baltimore, MD 21218, U.S.A. \and INAF-Istituto di Radio
  Astronomia, via P. Gobetti 101, I-40129 Bologna, Italy} \date{}
 
\abstract {From an optical spectroscopic survey of 3CR radio galaxies with
  $z<0.3$, we discovered a new spectroscopic class of powerful radio-loud
  AGN. The defining characteristics of these galaxies are that compared
  with radio galaxies of similar radio luminosity they have: a \oiiihb\
  ratio of $\sim$ 0.5, indicative of an extremely low level of gas excitation;
  a large deficit of \oiii\ emission and radio core power.  We interpret these
  objects as relic AGN, i.e. sources that experienced a large drop in their
  level of nuclear activity, causing a decrease in their nuclear and line
  luminosity. This class opens a novel approach to investigating lifetimes and
  duty cycles of AGN.}  

 \titlerunning{Discovery of the new spectroscopic class of relic
  radio galaxies} \authorrunning{A. Capetti et al.}

   \maketitle

\section{Introduction}
\label{introduction}

Optical spectroscopy has played a major role in enhancing our understanding of
active galactic nuclei (AGN).  \citet{heckman80} and \citet{baldwin81}
demonstrated that optical lines can be used as tools to classify in general
emission-line objects, particularly AGN. Diagnostic diagrams comparing emission
line ratios can distinguish H~II regions from gas clouds ionized by nuclear
activity \citep{veilleux87}. Moreover, AGN can be separated into Seyferts and
Low Ionization Nuclear Emission-line Regions \citep[LINERs,][]{heckman80}
since they form separate branches in the diagnostic diagrams
\citep{kewley06}.

We performed an optical spectroscopic survey of the 113 radio galaxies (RG)
belonging to the 3CR sample and with $z<$ 0.3 \citep{spinrad85}, using the
Telescopio Nazionale Galileo \citep{buttiglione09,buttiglione11}.  Most RGs
belong to two main spectroscopic classes, those of high and low excitation
galaxies (HEG and LEG respectively, originally introduced by
\citealt{laing94}, the analogous to Seyfert and LINER for radio-loud
AGN). In \citet{buttiglione10}, we also reported the discovery of a new class,
characterized by an extremely low level of gas excitation whose properties are
discussed in this Letter.

We adopt the following cosmological parameters: $H_o = 71$ km s$^{-1}$
Mpc$^{-1}$, $\Omega_{\Lambda} = 0.73$, and $\Omega_m = 0.27$.

\section{A new spectroscopic class: `Relic' radio galaxies}
\label{sect1}

\begin{figure*}[tp]
  \centerline{ \psfig{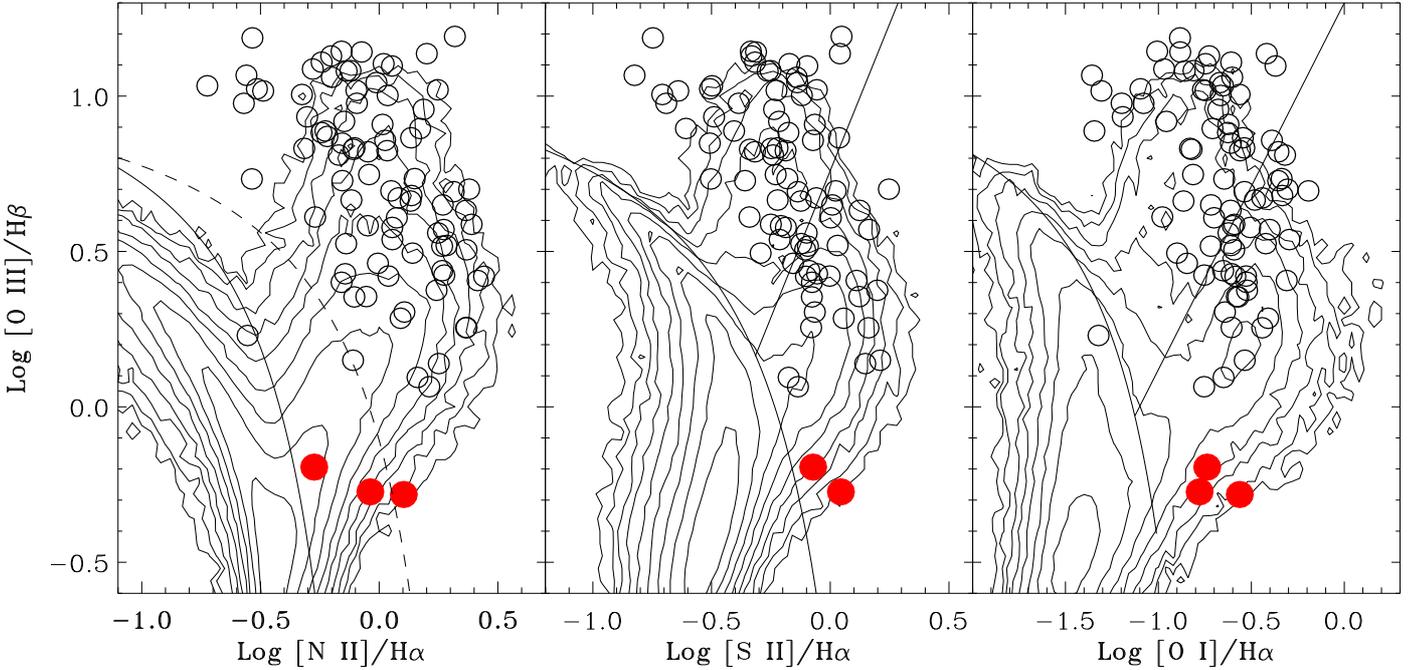}}
  \caption{\label{diag} Contours: density of SDSS emission-line galaxies in
    the optical diagnostic planes (adapted from \citealt{kewley06}). Galaxies
    below the curved solid lines are star-forming galaxies. In the left
    diagram, the dashed curve marks the transition from composite galaxies to
    AGN; in the middle and right panels, the oblique line separates LINERs from
    Seyferts. The circles are the 3CR sources. These diagrams reveal the
    existence of a class of extremely low excitation galaxies, i.e. the three
    filled red circles corresponding to the objects with the lowest
    \oiiihb\ ratios.}  \end{figure*}

Fig. \ref{diag} shows the location of the 3CR sources in the emission-line
diagnostic diagrams. They all fall in the area populated by AGN, with only one
exception represented by an object with emission line ratios typical of
star-forming galaxies. Their \oiiihb\ ratios are in the range 1.3 - 16.

Three galaxies (namely 3C~028, 3C~314.1, and 3C~348, shown in Fig. \ref{3c348}
as example\footnote{Since the [S~II] doublet is not covered by our spectrum,
  3C~348 is not represented in the central panel of Fig. \ref{diag}, where,
  for the same reason, the star-forming galaxy discussed above is also
  missing.})  stand out for a lower ratio, \oiiihb$\sim$ 0.5. For this reason,
in \citet{buttiglione10} we refer to this class as `extremely low excitation
galaxies' (ELEG). They are well separated from the rest of the 3CR sample and
fall in a region scarcely populated by Sloan Digital Sky Survey (SDSS,
\citealt{kewley06}) emission-line sources.  Considering first their [S
II]/\Ha\ and [O I]/\Ha\ ratios, they are located in the region of the
diagnostic diagram occupied by AGN. In the \oiiihb\ versus [N II]/\Ha\ plane,
they straddle the boundary separating AGN from composite galaxies, which are
possibly objects with a contribution from young stars and AGN to the emission
lines.

The ELEG are also distinctive because of their radio properties (see
Table. \ref{ccd}): 3C~028 and 3C~314.1 do not appear to contain a radio-core
at the level of 0.2 and 1.0 mJy respectively (see \citealt{giovannini88} and
references therein), while a 10 mJy radio core is detected in 3C~348
\citep{morganti93}.  The ratio of core (at 5 GHz) to total (at 178 MHz) radio
emission for these sources is in the range F$_{\rm core}$/F$_{\rm tot}
\lesssim 10^{-5} - 10^{-4}$, compared to an average ratio of $\sim 1.6 \times
10^{-3}$ for RGs of similar luminosity. They also have a \oiii\ line
luminosity that is 10 to 1000 times lower than RG of similar radio power (see
Fig. \ref{zoom}).\footnote{As noted above, two ELEG lie in Fig. \ref{diag}
  among the composite galaxies, generally thought to be objects with a mixed
  contribution from young stars and AGN to the emission lines. However, if
  this were the case for the ELEG, they should exhibit a line {\sl excess}
  with respect to RG of similar radio luminosity, in contrast to what is
  observed.}

\begin{table} 
\begin{center} 
\caption{Properties of the relic
      radio galaxies} \label{ccd} \begin{tabular}{l | c c c c} \hline \hline
      Name  & z & L \oiii\ & L$_{\rm core}$ & L$_{\rm 178}$ \\
      &   & \ergs & \ergsHz & \ergsHz \\
      \hline
      3C~028   & 0.195   & 40.96  & $<$29.33  & 34.24 \\
      3C~314.1 & 0.120   & 39.69  & $<$29.56  & 33.59 \\
      3C~348   & 0.154   & 40.41  &    30.80  & 35.35 \\
      \hline 
\end{tabular} 
\end{center} 
\end{table} 

On the basis of these findings, we interpret these objects as {\sl relic} AGN
in which the nuclear activity is currently switched-off (or substantially
weaker than its long-term average level). The quantities considered
(radio-core, emission-line luminosity and ratios, and total radio emission)
respond to changes in the AGN activity level on different timescales.  This is
because they are produced on scales of sub-parsec for the radio-core, to kpc
for the line emission produced in the narrow line region (NLR), to hundreds of
kpc for the extended radio-emission. A drop in the activity level is rapidly
followed by a decrease in both the radio core flux and the flux of ionizing
photons. This in turn causes the cooling of the NLR, leading to a relative low
level of \oiii$\lambda$5007 emission and a decrease in the \oiiihb\ ratio, as
described in more detail in the next section. In contrast, the luminosity
of the extended radio emission remains essentially unchanged over very long
timescales, accounting for the observed properties of ELEG.

\section{Spectroscopic time evolution of relic RG}
\label{sect2}

\citet{binette87} investigated the time evolution of the properties of a
photoionized cloud following a sharp decrease in the strength of the ionizing
photon field. They considered a plane-parallel cloud, in the low density
limit, and an instantaneous drop in the intensity of the nuclear emission by a
factor of 100. Their analysis found a very rapid quenching of the \oiii\
line. This is due to the high efficiency of the charge exchange reaction,
O$^{\,+\,2}$ + H$^{\,0}$ $\longrightarrow$ O$^{\,+}$ + H$^{\,+}$, which
proceeds rapidly as soon as the gas begins to recombine and the fraction of
neutral hydrogen increases.  They estimated that the decay time of the \oiii\
(the time after which the line intensity has decreased by a factor 1/$e$) is
$t_d({\rm [O~III]}) \sim 2000 \, n_e^{-1}$ years, where $n_e$ is the
electronic density.  Since the decay time of the \Hb\ line is substantially
longer, $t_d ({\rm H}\beta) \sim 130000 \, n_e^{-1}$ years, the \oiiihb\ ratio
decreases with time. Starting from an initial value of \oiiihb$\sim$15, after
$\sim 4000 \, n_e^{-1}$ years the ratio has decreased to $\sim$ 0.5 (the value
observed in ELEG).

\begin{figure*}[htbp] \centerline{
    \psfig{figure=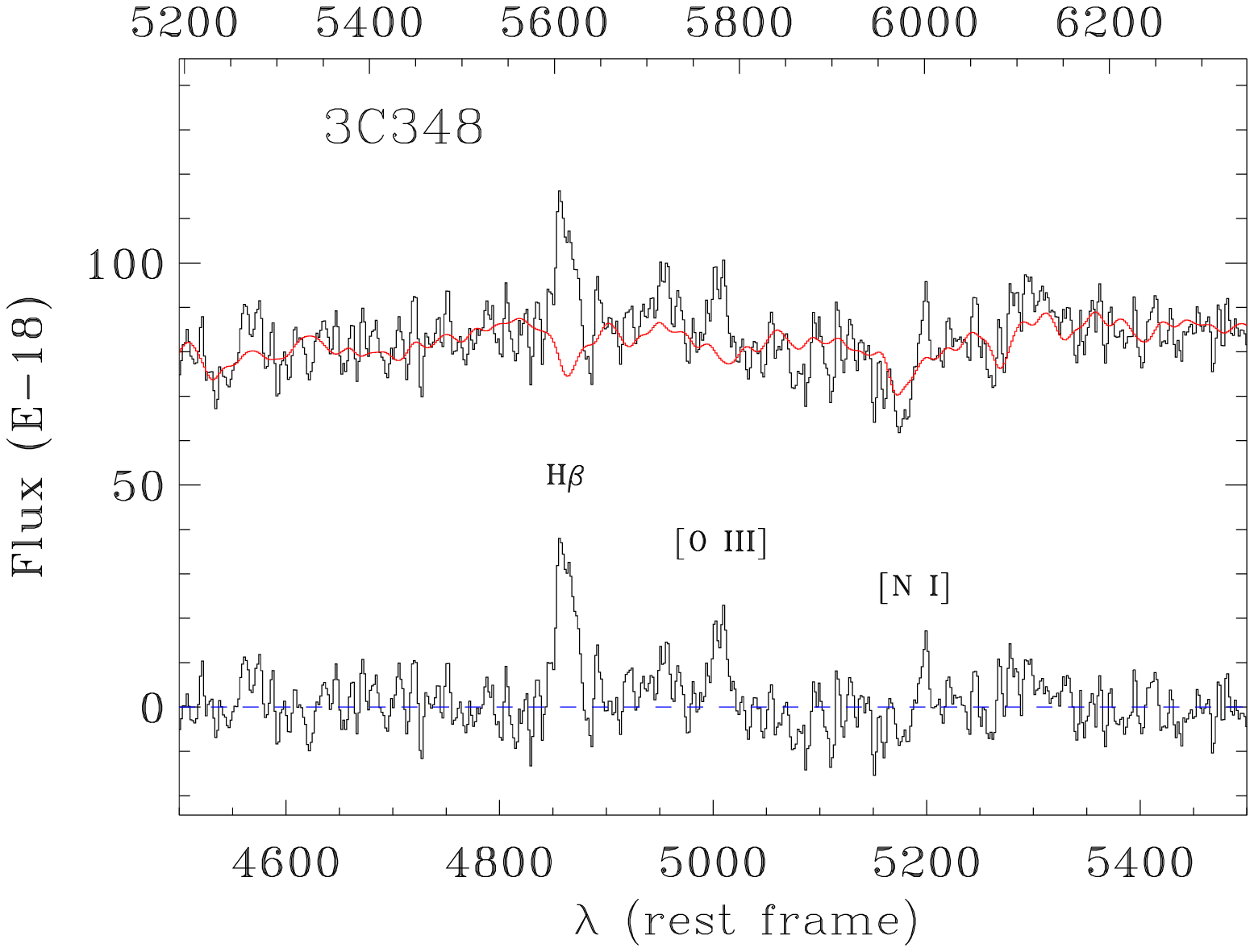,angle=0,width=0.5\linewidth}
    \psfig{figure=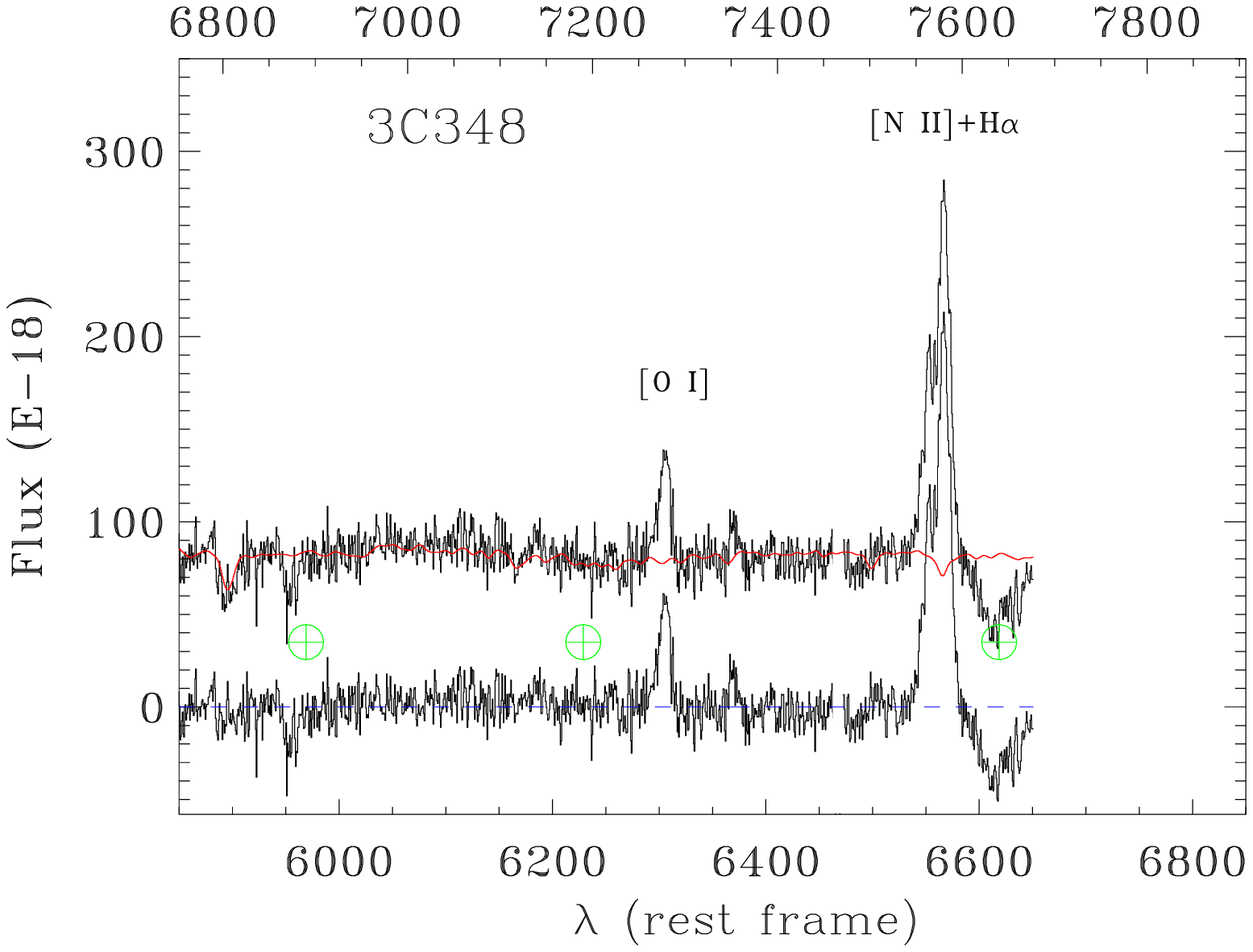,angle=0,width=0.5\linewidth}}
  \caption{\label{3c348} TNG spectrum of 3C348, an example of a relic RG. The
    left and right panels are centered on the \Hb\ and \Ha\ spectral regions,
    respectively. The original spectra are indicated by the upper solid lines
    with superimposed we show the best-fit stellar populations model used to
    subtract the starlight; the lower solid lines are the residual
    spectra. The flux is given in $10^{-18}$ erg cm$^{-2}$ s$^{-1}$ \AA$^{-1}$,
    while the wavelength is in \AA, in the rest frame in the lower axis and in
    the observed frame in the axis above. The three main telluric absorption
    bands are indicated as circles with a cross inside.}  \end{figure*}

All relevant timescales depend on density. The density tracers corresponding
to the brightest emission lines are the [O II]$\lambda\lambda$3726,3729 and [S
II]$\lambda\lambda$6716,6731 doublets. The individual measurement of the [O
II] lines is irreparably compromised by the intrinsic line widths in the NLR,
leaving one only with the possibility of exploiting the [S II] doublet alone.
Unfortunately, from our spectra we cannot obtain an accurate estimate of this
parameter. We have no measurements of the [S II] for 3C~348 (not covered by
the TNG data, see the right panel in Fig.  \ref{3c348}), while in our spectrum
of 3C~028 the [S II]$\lambda$6731 line falls in a CCD defect. Only for
3C~314.1 are we able to measure a ratio [S II]$\lambda$6716/[S
II]$\lambda$6731 = 1.6 $\pm$ 0.3, which is indicative only of a low density
($n_e \lesssim 3 \times 10^2$ cm$^{-3}$) regime \citep{osterbrock89}.

An additional factor to consider in the temporal evolution is the light travel
time from the nucleus to the NLR clouds. The light travel time from the
nucleus to the edge of the 2$\arcsec$ slit is $\sim 10^4$ years. This must be
considered as an upper limit to the response of the portion of the NLR probed
by our observations to a change in the nuclear flux since emission lines of RG
are usually highly concentrated around the nucleus \citep[e.g.][]{tremblay09}.
Nonetheless, considering the uncertainties related to the actual distribution
and density of ionized gas, it can be envisaged that the light travel effects
might play a key role in the evolution of the observable properties of the
NLR.

\begin{figure}[htbp] 
\centerline{
    \psfig{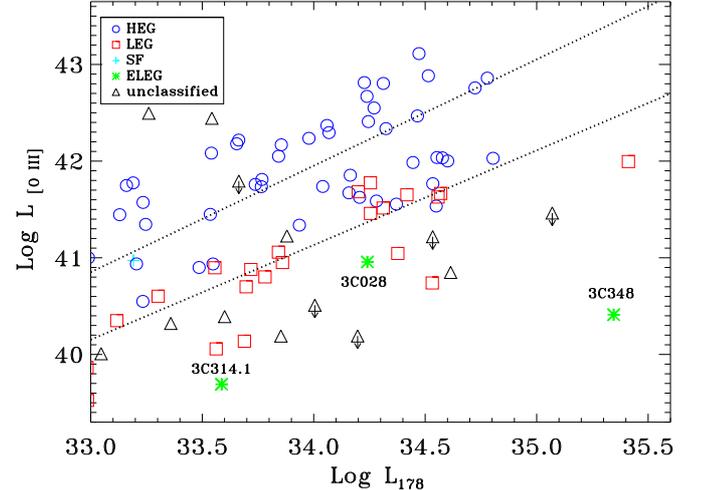}}
  \caption{\label{zoom} \oiii\ vs. total radio luminosity at 178 MHz (in
    units of \ergs\ and \ergsHz\ respectively) for the 3CR sources with
    z$<$0.3 (limiting for clarity to its bright luminosity end). HEG are
    marked with circles, LEG with squares. The green asterisks are ELEG,
    i.e. the relic RG. The triangles are objects that could not be classified
    spectroscopically. The dotted lines are the best-fit linear relation
    between $L_{\rm[O III]}$ and $L_{178}$ obtained for HEG and LEG
    separately.}
\end{figure}

\section{Discussion and conclusions}
\label{summary}

The energy carried by the jets of radio-loud AGN has a profound impact on the
evolution of their hosts and the energy balance of the intracluster medium
(e.g. \citealt{fabian03,croton06}). To ascertain the effects of AGN feedback,
it is necessary to measure not only the jet power, but also the timescales and
frequency of the energy release, i.e. measure AGN lifetimes and duty
cycles. Several approaches, based purely on radio properties, have been used
previously (e.g. \citealt{alexander87,shabala08,bird08}) to investigate these
issues. These analyses have led to evidence \citep{giovannini88,parma07} of
relic (fossil) RG based on the low core dominance or very steep spectral
indices (a manifestation of the energy losses of the radio-emitting
relativistic electrons) of some systems.  Using radio observations it is
possible to derive model-dependent estimates of the RG lifetimes or, in case
of fossil RG, the time elapsed since the `death' of the active nucleus,
$\tau_{\rm D}$.

Relic RGs, distinguished by their optical spectroscopic properties, open a new
and complementary path to exploring the lifetime and duty cycles of radio
galaxies. The fraction of relics in the RG population is given by the ratio of
the duration of the relic phase to the RG lifetime. In principle, both
parameters can be derived once all RG in a complete sample are
spectroscopically classified, a situation that is (almost) met by our 3CR
spectroscopic survey.

Considering the range of extended luminosity where relic RG are observed, Log
L$_{178} > 33$ \ergsHz, there are 83 objects in the 3CR with $z<0.3$, three of
which are relics, one has a star-forming-like spectrum, 46 are HEG, and 20 are
LEG. This leads to the fraction of relics being 3/83.  This fraction can
increase substantially depending on the nature of the 13 objects in the sample
that could not be classified spectroscopically. In particular, as explained in
more detail below, some of them can be considered as `candidate relics'. A
second, less important, uncertainty in the relic fraction would also exist if
HEG only can evolve into relic systems.

  The duration of the relic phase could in principle be estimated using the
  spectroscopic evolution, by comparing the temporal changes in line ratios
  with those predicted by photoionization models.  Furthermore, the intensity
  of the various lines can be compared with those of RG of similar extended
  luminosity, i.e. with the relic progenitors.  This approach requires a
  direct measurement of the NLR density, the parameter that drives its
  temporal evolution. Unfortunately, the density in the relic galaxies is
  essentially unconstrained by our observations as we are unable to measure
  the [S~II] doublet ratio in two of the ELEG, and it is poorly constrained
  for 3C~314.1.  However, future measurements should be easily attainable with
  dedicated observations as the intensity of the [S~II] doublet is similar to
  that of the \Ha\ line, whose flux was accurately measured in our spectra.

  Although the extended radio emission is, in this context, the structure
  responding on the longer timescale to the changes in the nuclear properties,
  the radio morphology of the three relics can provide useful insights into
  their evolution.  All of them can be classified as FR~II following the
  original definition of \citet{fanaroff74}, but they differ when examined in
  more detail:

  -- 3C~028 has well-defined twin jets linking the host to a double-lobed
  structure, with well-defined hot spots, but lacks a radio core
  \citep{feretti84}. There is no apparent brightness discontinuity in the jets
  on either side of the host, an indication that the drop in nuclear activity
  occurred recently.

  -- 3C~314.1 has a structure known as a ``fat double'' as it lacks jets and
  hot spots \citep{leahy91}.  This is an indication that the switching-off of
  the jets of this galaxy occurred at least $\sim 7 \times 10^5$ years ago,
  based on the light travel time to the edge of the radio source $\sim$ 250
  kpc from the nucleus.

  -- 3C~348 is considered to be a ``born again" RG based on its peculiar radio
  morphology \citep{gizani03}: two jets emerge from the core and propagate
  within a relaxed double-lobed, steep-spectrum structure. On the basis of the
  size of the inner radio structure embedded in the FR~II fossil ($\sim$ 200
  kpc), the relic phase started at least $3\times 10^5$ years ago. Despite the
  presence of a restarted AGN, the relic line emission appears to be still the
  dominant component observed in this galaxy; this implies that the current
  phase must be associated with a significantly lower level of nuclear
  luminosity than the earlier phase of activity.
 
  Evidence of the different evolutionary stages in the three sources is
  supported by comparing their \oiii\ luminosity with respect to HEG of
  similar radio luminosity (see Fig. \ref{zoom}): for 3C~028 this is lower by
  only a factor of $\lesssim$ 10, while 3C~314.1 and 3C~348 exhibit much
  larger deficits (a factor of 10$^2$ - 10$^3$) .

  An independent estimate of the relic phase duration, $\tau_{\rm D}$, can be
  obtained by {\sl directly} spatially mapping the change in state of the AGN
  using deeper optical spectroscopy of the off-nuclear emission line regions
  (e.g. \citealt{robinson87}). Owing to light travel effects, information
  about the drop in the nuclear emission might not have reached these outer
  regions. In a sort of reversed light-echo effect, the regions located at
  radii $r > c \, \tau_{\rm D}$ would still show the original high excitation
  state, characterized by a ratio \oiiihb\ $\sim 20$ times higher than in the
  nuclear regions. In practice, geometrical effects affect the estimate
  of $\tau_{\rm D}$, since the isochrones are paraboloids centered on the
  nucleus and the derived timescale depends on the orientation of the gas
  cloud with respect to the line of sight. This degeneracy can be broken when
  emission lines are detected on both sides of the nucleus.

  Finally, we consider in more detail the 13 spectroscopically unclassified
  sources (usually because the \oiii\ and/or the H$\beta$ lines are too faint
  to be measured).  Not all of them should be considered as potential
  additional relics since the uncertain classification is caused by a variety
  of factors\footnote{ For example, 3C~111 and 3C~445 are broad line RG, not
    classified because their broad Balmer lines hides completely the narrow
    components; 3C~132 is seen through a region of very high galactic
    absorption ($A_V \sim 4$); in 3C~346 the \Ha\ line coincides with a
    telluric band.}.  However, relative to RG of similar total radio
  luminosity nine unclassified sources also have a relatively low \oiii\
  luminosity (see Fig. \ref{zoom}) and we therefore consider them to be
  plausible `candidate relics'.  The importance of these objects is two-fold:
  first of all, only after a proper spectroscopic validation will it be
  possible to derive the ratio of the number of relic to active galaxies, and
  consequently the relative duration of the two phases; secondly, it is
  possible that these sources might be in a different evolutionary stage than
  the three relic galaxies discussed here.  A clearer characterization of the
  class of relic RG would require a study of these objects in greater detail
  from both the radio and spectroscopic point of view.

  Relics AGN are also likely to exist among radio-quiet objects; for example,
  NGC~5252, with its high-excitation extended NLR surrounding a nuclear region
  with a LINER spectrum, is a likely relic QSO
  \citep{goncalves98,capetti5252}. In the region of the spectroscopic diagrams
  typical of the 3CR relics there is indeed a substantial number of SDSS
  emission-line galaxies (see Fig. 1). However, the lack of radio
  diagnostics makes it identify genuine relics among the broad
  distribution of radio-quiet AGN in the diagnostic planes.  Furthermore,
  relics are likely to be rare in a flux-limited sample of emission line
  galaxies, particularly when their definition requires the detection of the
  short-lived \oiii\ line. As an alternative, one could rely on non-classical
  diagnostic diagrams, including lines originating from high excitation gas
  unaffected by the charge exchange reaction (e.g. HeII$\lambda$4686) that
  causes the extremely short-decay time for the [O~III] line. The temporal
  evolution within these same diagnostic diagrams can provide us with an
  estimate of the relic age. However, the likelihood of observing these
  effects will in general depend on both the gas density and (through light
  travel times) the size of the emitting region. In principle, extrapolating
  backward in time it will be possible to associate these sources with their
  active counterparts at the correct level of luminosity and thus estimate
  their lifetimes.

\bibliographystyle{aa}


\begin{thebibliography}{27}
\expandafter\ifx\csname natexlab\endcsname\relax\def\natexlab#1{#1}\fi

\bibitem[{{Alexander} \& {Leahy}(1987)}]{alexander87}
{Alexander}, P. \& {Leahy}, J.~P. 1987, MNRAS, 225, 1

\bibitem[{{Baldwin} {et~al.}(1981){Baldwin}, {Phillips}, \&
  {Terlevich}}]{baldwin81}
{Baldwin}, J.~A., {Phillips}, M.~M., \& {Terlevich}, R. 1981, \pasp, 93, 5

\bibitem[{{Binette} \& {Robinson}(1987)}]{binette87}
{Binette}, L. \& {Robinson}, A. 1987, \aap, 177, 11

\bibitem[{{Bird} {et~al.}(2008){Bird}, {Martini}, \& {Kaiser}}]{bird08}
{Bird}, J., {Martini}, P., \& {Kaiser}, C. 2008, \apj, 676, 147

\bibitem[{{Buttiglione} {et~al.}(2009){Buttiglione}, {Capetti}, {Celotti},
  {Axon}, {Chiaberge}, {Macchetto}, \& {Sparks}}]{buttiglione09}
{Buttiglione}, S., {Capetti}, A., {Celotti}, A., {et~al.} 2009, \aap, 495, 1033

\bibitem[{{Buttiglione} {et~al.}(2010{\natexlab{a}}){Buttiglione}, {Capetti},
  {Celotti}, {Axon}, {Chiaberge}, {Macchetto}, \& {Sparks}}]{buttiglione10}
{Buttiglione}, S., {Capetti}, A., {Celotti}, A., {et~al.} 2010,
  \aap, 509, A6

\bibitem[{{Buttiglione} {et~al.}(2010{\natexlab{b}}){Buttiglione}, {Capetti},
  {Celotti}, {Axon}, {Chiaberge}, {Macchetto}, \& {Sparks}}]{buttiglione11}
{Buttiglione}, S., {Capetti}, A., {Celotti}, A., {et~al.} 2011,
 \aap, 525, A28

\bibitem[{{Capetti} {et~al.}(2005){Capetti}, {Marconi}, {Macchetto}, \&
  {Axon}}]{capetti5252}
{Capetti}, A., {Marconi}, A., {Macchetto}, D., \& {Axon}, D. 2005, \aap, 431,
  465

\bibitem[{{Croton} {et~al.}(2006){Croton}, {Springel}, {White}, {De Lucia},
  {Frenk}, {Gao}, {Jenkins}, {Kauffmann}, {Navarro}, \& {Yoshida}}]{croton06}
{Croton}, D.~J., {Springel}, V., {White}, S.~D.~M., {et~al.} 2006, MNRAS, 365,
  11

\bibitem[{{Fabian} {et~al.}(2003){Fabian}, {Sanders}, {Allen}, {Crawford},
  {Iwasawa}, {Johnstone}, {Schmidt}, \& {Taylor}}]{fabian03}
{Fabian}, A.~C., {Sanders}, J.~S., {Allen}, S.~W., {et~al.} 2003, \mnras, 344,
  L43

\bibitem[{{Fanaroff} \& {Riley}(1974)}]{fanaroff74}
{Fanaroff}, B.~L. \& {Riley}, J.~M. 1974, MNRAS, 167, 31P

\bibitem[{{Feretti} {et~al.}(1984){Feretti}, {Gioia}, {Giovannini},
  {Gregorini}, \& {Padrielli}}]{feretti84}
{Feretti}, L., {Gioia}, I.~M., {Giovannini}, G., {Gregorini}, L., \&
  {Padrielli}, L. 1984, \aap, 139, 50

\bibitem[{{Giovannini} {et~al.}(1988){Giovannini}, {Feretti}, {Gregorini}, \&
  {Parma}}]{giovannini88}
{Giovannini}, G., {Feretti}, L., {Gregorini}, L., \& {Parma}, P. 1988, \aap,
  199, 73

\bibitem[{{Gizani} \& {Leahy}(2003)}]{gizani03}
{Gizani}, N.~A.~B. \& {Leahy}, J.~P. 2003, MNRAS, 342, 399

\bibitem[{{Goncalves} {et~al.}(1998){Goncalves}, {Veron}, \&
  {Veron-Cetty}}]{goncalves98}
{Goncalves}, A.~C., {Veron}, P., \& {Veron-Cetty}, M. 1998, \aap, 333, 877

\bibitem[{{Heckman}(1980)}]{heckman80}
{Heckman}, T.~M. 1980, \aap, 87, 152

\bibitem[{{Kewley} {et~al.}(2006){Kewley}, {Groves}, {Kauffmann}, \&
  {Heckman}}]{kewley06}
{Kewley}, L.~J., {Groves}, B., {Kauffmann}, G., \& {Heckman}, T. 2006, \mnras,
  372, 961

\bibitem[{{Laing} {et~al.}(1994){Laing}, {Jenkins}, {Wall}, \&
  {Unger}}]{laing94}
{Laing}, R.~A., {Jenkins}, C.~R., {Wall}, J.~V., \& {Unger}, S.~W. 1994, in
  Astronomical Society of the Pacific Conference Series, Vol.~54, The Physics
  of Active Galaxies, ed. G.~V. {Bicknell}, M.~A. {Dopita}, \& P.~J. {Quinn},
  201

\bibitem[{{Leahy} \& {Perley}(1991)}]{leahy91}
{Leahy}, J.~P. \& {Perley}, R.~A. 1991, \aj, 102, 537

\bibitem[{{Morganti} {et~al.}(1993){Morganti}, {Killeen}, \&
  {Tadhunter}}]{morganti93}
{Morganti}, R., {Killeen}, N.~E.~B., \& {Tadhunter}, C.~N. 1993, MNRAS, 263,
  1023

\bibitem[{{Osterbrock}(1989)}]{osterbrock89}
{Osterbrock}, D.~E. 1989, Astrophysics of Gaseous Nebulae and Active Galactic
  Nuclei (Mill Valley: Univ. Sci. Books)

\bibitem[{{Parma} {et~al.}(2007){Parma}, {Murgia}, {de Ruiter}, {Fanti},
  {Mack}, \& {Govoni}}]{parma07}
{Parma}, P., {Murgia}, M., {de Ruiter}, H.~R., {et~al.} 2007, \aap, 470, 875

\bibitem[{{Robinson} {et~al.}(1987){Robinson}, {Binette}, {Fosbury}, \&
  {Tadhunter}}]{robinson87}
{Robinson}, A., {Binette}, L., {Fosbury}, R.~A.~E., \& {Tadhunter}, C.~N. 1987,
  MNRAS, 227, 97

\bibitem[{{Shabala} {et~al.}(2008){Shabala}, {Ash}, {Alexander}, \&
  {Riley}}]{shabala08}
{Shabala}, S.~S., {Ash}, S., {Alexander}, P., \& {Riley}, J.~M. 2008, \mnras,
  388, 625

\bibitem[{{Spinrad} {et~al.}(1985){Spinrad}, {Marr}, {Aguilar}, \&
  {Djorgovski}}]{spinrad85}
{Spinrad}, H., {Marr}, J., {Aguilar}, L., \& {Djorgovski}, S. 1985, \pasp, 97,
  932

\bibitem[{{Tremblay} {et~al.}(2009){Tremblay}, {Chiaberge}, {Sparks}, {Baum},
  {Allen}, {Axon}, {Capetti}, {Floyd}, {Macchetto}, {Miley}, {Noel-Storr},
  {O'Dea}, {Perlman}, \& {Quillen}}]{tremblay09}
{Tremblay}, G.~R., {Chiaberge}, M., {Sparks}, W.~B., {et~al.} 2009, \apjs, 183,
  278

\bibitem[{{Veilleux} \& {Osterbrock}(1987)}]{veilleux87}
{Veilleux}, S. \& {Osterbrock}, D.~E. 1987, \apjs, 63, 295

\end{thebibliography}

\end{document}